\documentclass{WileyMSP-template}
\usepackage[utf8]{inputenc}
\usepackage{amsmath}
\usepackage{upgreek}

\begin{document}


\title{Quantum diamond microscopy of individual vaterite microspheres containing magnetite nanoparticles}

\maketitle


\author{Mona Jani*}
\author{Hani Barhum}
\author{Janis Alnis}
\author{Mohammad Attrash}
\author{Tamara Amro}
\author{Nir Bar-Gill}
\author{Toms Salgals}
\author{Pavel Ginzburg}
\author{Ilja Fescenko*}

\begin{affiliations}
M. Jani, I. Fescenko\\
Laser Center, University of Latvia, Riga LV-1004, Latvia\\
Email: ilja.fescenko@lu.lv, mona.jani@lu.lv

H. Barhum, M. Attrash\\
Triangle Regional Research and Development Center, Kfar Qara’ 3007500, Israel\\
and School of Electrical Engineering, Tel Aviv University, Ramat Aviv, Tel Aviv 69978, Israel

J. Alnis\\
Institute of Atomic Physics and Spectroscopy, University of Latvia, Riga LV-1004, Latvia

T. Amro, N. Bar-Gill\\
The Racah Institute of Physics, The Hebrew University of Jerusalem, Jerusalem 91904, Israel

P. Ginzburg\\
School of Electrical Engineering, Tel Aviv University, Ramat Aviv, Tel Aviv 69978, Israel

T. Salgals\\
Institute of Telecommunications, Riga Technical University, 1048 Riga, Latvia

\end{affiliations}



\begin{abstract}

Biocompatible vaterite microspheres, renowned for their porous structure, are promising carriers for magnetic nanoparticles (MNPs) in biomedical applications such as targeted drug delivery and diagnostic imaging. Precise control over the magnetic moment of individual microspheres is crucial for these applications. This study employs widefield quantum diamond microscopy to map the stray magnetic fields of individual vaterite microspheres (3-$10~\upmu$m) loaded with Fe$_3$O$_4$ MNPs of varying sizes (5~nm, 10~nm, and 20~nm). By analyzing over 35 microspheres under a 222 mT external magnetizing field, we measured peak-to-peak stray field amplitudes of $41 \pm 1~\upmu$T for 5~nm and 10~nm superparamagnetic MNPs, reflecting their comparable magnetic response, and $12 \pm 1~\upmu$T for 20~nm ferrimagnetic MNPs, due to distinct magnetization behavior. Finite-element simulations confirm variations in MNP distribution and magnetization uniformity within the vaterite matrix, with each microsphere encapsulating thousands of MNPs to generate its magnetization. This high-resolution magnetic imaging approach yields critical insights into MNP-loaded vaterite, enabling optimized synthesis and magnetically controlled systems for precision therapies and diagnostics.

\end{abstract}


\section{Introduction}

Vaterite particles, a naturally occurring mineral form of polycrystalline calcium carbonate (CaCO$_3$), are both biocompatible and biodegradable, making them promising candidates for targeted drug delivery~\cite{Vikulina2018,Demina2024,Huang2024,Ferreira2023,Bahrom2019,Harpaz2023}. Their mesoporous spherical structure allows the embedding of diverse materials, such as metal nanoparticles or bio-macromolecules~\cite{Parakhonskiy2014,Azarian2024, Barhum2024a,Barhum2024b}. The high loading capacity of these particles increases their potential for targeted and sustained drug delivery~\cite{Xia2020}. Additionally, upon exposure to solvents, vaterite undergoes a phase transition to calcite, facilitating the controlled release of the embedded materials~\cite{Gilad2025}. When vaterite particles are loaded with magnetic nanoparticles, their applicability extends to medical imaging, targeted drug delivery and hyperthermia. MNPs such as magnetite (Fe$_3$O$_4$) embedded in vaterite microspheres enable controlled movement and displacement using magnetic fields, making them suitable for precision drug delivery. The high porosity of vaterite microspheres allows it to host significant amounts of Fe$_3$O$_4$, with a reported loading capacity of 11~\text{mass \%}~\cite{Choukrani2020}. Research by N. Markina \textit{et al.}~\cite{Markina2017} and B. V. Parakhonskiy \textit{et al.}~\cite{Parakhonskiy2019} have demonstrated their potential in magnetic separation, movement, and displacement for biomolecule detection. MNP-embedded vaterite particles with highly uniform composition and magnetic properties would greatly benefit these applications. Uniformly magnetized Fe$_3$O$_4$-vaterite particles would allow more precise control over drug delivery and hyperthermia therapy~\cite{Richards2025}. However, commercially available tools for quantitative magnetometry at the single-particle level with micrometer resolution are rather complex and rare. Advanced characterization techniques, while potentially useful, are often prohibitively expensive, require complex maintenance, or are simply nonexistent. To enable rapid screening and optimization of nanoparticles, customized tools need to be developed.

The magnetic properties of Fe$_3$O$_4$ MNPs have been extensively studied using various techniques, including vibrating-sample magnetometry~\cite{Frandsen2021}, superconducting quantum interference device (SQUID)~\cite{Hajalilou2023}, and M\"{o}ssbauer spectroscopy~\cite{Johnson2016}. These methods provide valuable insights into the bulk magnetic behavior of MNP ensembles. However, the magnetic field of MNPs within an individual vaterite particle remains largely unexplored. While techniques like micro-SQUID magnetometry~\cite{Adolphi2010}, scanning magnetic microscopy~\cite{Araujo2020}, Brillouin light scattering~\cite{Sebastian2015}, Lorentz electron microscopy~\cite{Kirk1999}, and magnetic force microscopy~\cite{Sievers2012} offer the potential for single-particle magnetic imaging, they often face limitations in sensitivity for detecting the weak magnetic fields of individual particles or require specific, sometimes challenging, operating conditions.

Recently, widefield quantum diamond microscopy (QDM) has been developed to map magnetic fields with high sensitivity and sub-micron spatial resolution~\cite{Levine2019, Fescenko2019}. This technique relies on the optical detection of electron spin resonance from Nitrogen-Vacancy (NV) centers in diamond. The NV centers in a diamond have a well-characterized electronic structure that is sensitive to magnetic fields and can be examined using Optically Detected Magnetic Resonance (ODMR) spectroscopy under ambient conditions~\cite{Berzins2023,Jani2025}. By depositing individual particles directly onto the diamond surface, optical spatial resolutions as fine as $400$ nm can be achieved~\cite{Fescenko2019}. QDM has been used to map stray magnetic fields from ferrimagnetic~\cite{LeSage2013}, superparamagnetic~\cite{Richards2025}, and paramagnetic MNPs~\cite{Lamichhane2024a,Lamichhane2024b,Lamichhane2023}. While small magnetite nanoparticles are typically superparamagnetic at room temperature and generate strong signals near NV centers, this does not apply to larger Fe$_3$O$_4$-functionalized vaterite microspheres, which produce weaker stray fields. In these microspheres, most MNPs are expected to be micrometers away from the NV layer. Since the dipole magnetic field rapidly decays with distance, studying the magnetic field associated with MNPs within individual vaterite particles remains both a challenge and an opportunity to explore more strategies for magnetically responsive platforms.
        	
In this study, we used widefield QDM to investigate the stray magnetic fields ($B_{str}$) of Fe$_3$O$_4$ MNPs (5-nm, 10-nm, and 20-nm) embedded within vaterite microspheres. Applying an external magnetizing field of 222 mT, we mapped the stray fields and measured the peak-to-peak (PP) amplitudes of the magnetic signals from $> 35$ individual vaterite microspheres, ranging in size from 5 to $10~\upmu$m. These measurements allowed us to correlate the observed stray fields with both the MNP and vaterite microsphere sizes, independently estimated using optical microscopy. Furthermore, we compared the measured magnetic maps with simulated maps generated by finite-element analysis of homogeneously magnetized microspheres. This comparison served to assess the uniformity of magnetization within the vaterite microspheres due to the embedded MNPs. Our study demonstrates a novel approach for investigating the magnetic fields of Fe$_3$O$_4$ MNPs in biogenic minerals such as vaterite, offering new insights into their magnetic properties and potential applications.

\section{Experimental methods}

\threesubsection{Synthesis of Fe$_3$O$_4$ loaded vaterite microspheres}

The Fe$_3$O$_4$ MNPs were synthesized via the co-precipitation method to achieve specific sizes (10±2 nm, and 20±2 nm) while preserving their ferromagnetic properties we used ultrasonication assisted synthesis~\cite{MartinezMera2007}. The particle sizes were confirmed using transmission electron microscopy (TEM). Additionally, 5, 10, and 20~nm particles with ultra low dispersion were purchased from Merk dispersed in toluene and from NN Crystal Ltd dispersed in DIW having PEG on its surface. The magnetization ($M$) of the Fe$_3$O$_4$ MNPs ranged from 1.55-$2.59 \times 10^6$~A/m for 5 nm particles~\cite{Hadadian2022}, $3.27-3.64  \times 10^6$~A/m for 10 nm particles~\cite{Nkurikiyimfura2020}, and $3.8-4.7  \times 10^5$~A/m for 20 nm particles~\cite{Hadadian2022}. The 5~nm and 10~nm Fe$_3$O$_4$ MNPs exhibited strong superparamagnetic behavior, whereas the 20~nm particles displayed ferrimagnetic properties with moderate magnetic susceptibility.

\begin{figure*}[!t]
  \begin{center}
  \includegraphics[width=0.9\linewidth]{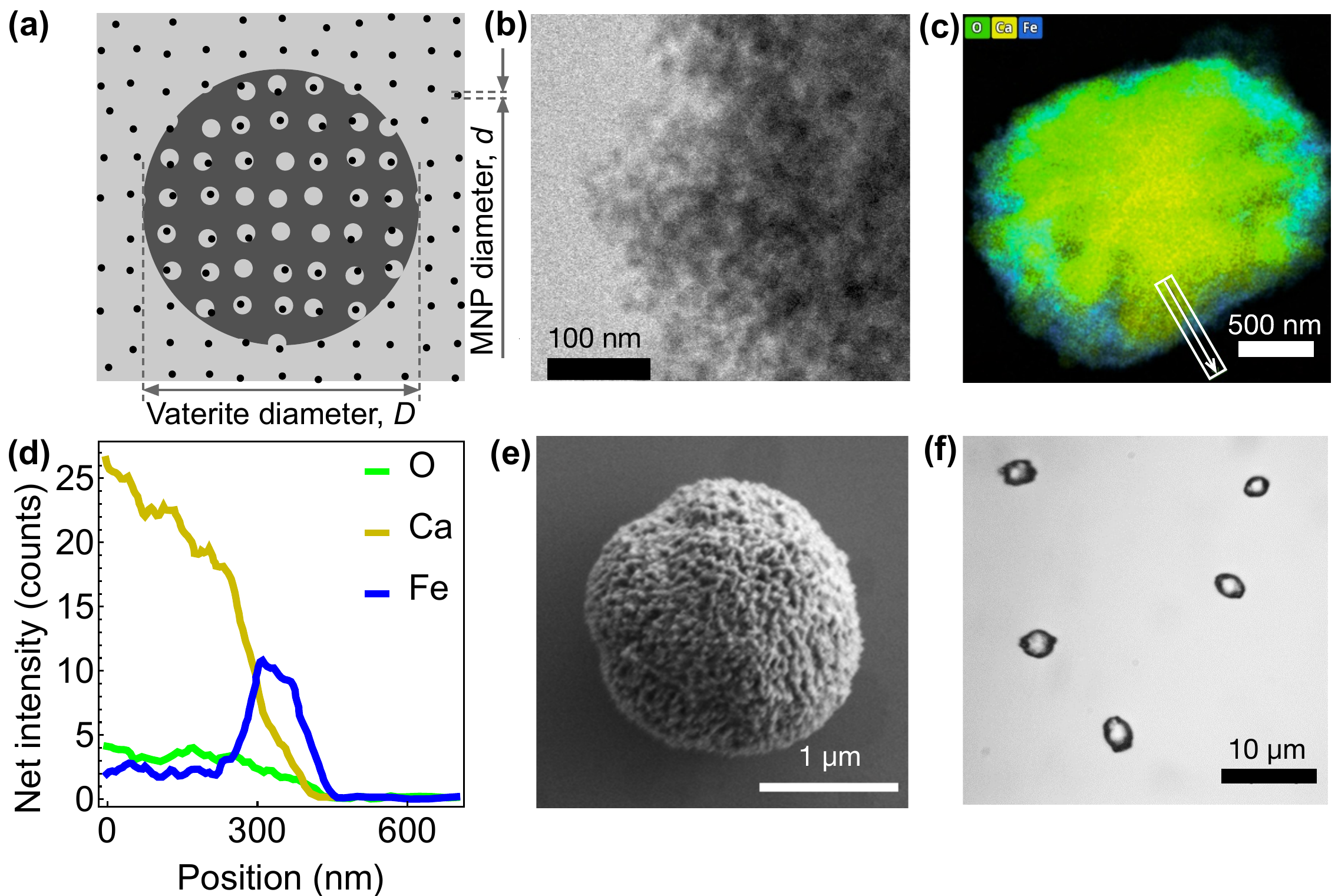}
  \end{center}
  \caption{\textbf{Characterization of vaterite microspheres loaded with Fe$_3$O$_4$ magnetic nanoparticles.} (a) Schematic illustration of a porous vaterite microsphere (dark gray) loaded with Fe$_3$O$_4$ MNPs (black dots) dispersed within its structure from the surrounding suspension (gray). (b) High-resolution TEM image of the edge of a vaterite microsphere, revealing the presence of 10 nm Fe$_3$O$_4$ MNPs. (c) Energy-dispersive X-ray spectroscopy (EDS) elemental mapping of an entire magnetic vaterite microparticle. (d) Quantitative elemental analysis of the microsphere’s surface along the marked region indicated in (c), displaying the relative abundance of O, Ca, and Fe. (e) Scanning electron microscopy (SEM) image of a vaterite microsphere loaded with 20 nm Fe$_3$O$_4$ MNPs, highlighting its surface morphology. (f) Optical microscopy images of individual vaterite microspheres drop-casted onto a glass coverslip, observed using a 100× oil immersion objective.}
  \label{fig:Fig1}
\end{figure*}

Vaterite particles were synthesized using a conventional co-precipitation reaction procedure with slight modifications. Briefly, a stock solution of 1 M calcium chloride (CaCl$_2$) and 1 M sodium carbonate (Na$_2$CO$_3$) solutions were mixed after dilution to 0.3~M with and without addition of poly-styrenesulfonate 70k, at 1 mg/mL achieving more polar surface and larger particles. Mixing of the solutions stopped after 1 min, then the particles were washed with DI water followed by three EtOH washes, the sample was then dried or kept in EtOH for storing. Achieving smaller particles was through using controlled synthesis in ethylene glycol 85\% as was shown previously~\cite{Bahrom2019}. The characterization of the resulting CaCO$_3$ particles, including their size and porosity, is described in detail by Barhum \textit{et al.}~\cite{Bahrom2019}. The addition of MNPs to vaterite involved 15 minutes of vortexing, followed by 1~hour of sonication, then another 15 minutes of vortexing and an additional 1~hour of sonication. Then, the particle solutions were left for a few minutes to rest and the remaining MNPs in solution were removed, and EtOH was added then washed three times through centrifugation. The pore size, surface charges and channel structure of the vaterite particles influenced the incorporation and distribution of Fe$_3$O$_4$ MNPs within them. \textbf{Figure~\ref{fig:Fig1}(a)} depicts a schematic representation of the process by which a spherical, porous vaterite microsphere (dark gray) is loaded with Fe$_3$O$_4$ MNPs (black dots) from a suspension (gray). Prior to loading, the MNP, and vaterite suspensions were sonicated for 1~minute to ensure uniform dispersion. Subsequently, the particles were rinsed with ethanol to remove residual water and dried at 60$^{\circ}$C for 3 hours under air convection. The dried Fe$_3$O$_4$-vaterite powder (approximately 90~mg) was stored in polyethylene test tubes under ambient conditions. For magnetic imaging, the Fe$_3$O$_4$-vaterite microspheres were resuspended in 99.8\% pure ethanol, vortexed for a few seconds, and drop-cast onto the surface of a diamond sensor.

High-resolution transmission electron microscopy imaging of the edge of a vaterite microsphere loaded with 10~nm Fe$_3$O$_4$ MNPs is shown in \textbf{Figure~\ref{fig:Fig1}(b)}. The MNPs appear as dark spots within the microsphere, with an estimated density of $50-90$~MNPs per 200~nm$^2$ area. Energy-dispersive X-ray spectroscopy (EDS) elemental mapping (\textbf{Figure~\ref{fig:Fig1}(c)}) confirms the presence of iron (Fe) and oxygen (O) within the microsphere, alongside calcium (Ca) from the vaterite matrix, with the iron distribution appearing concentrated closer to the surface. An EDS line scan along the path marked by the black line in Figure~\ref{fig:Fig1}(c) is presented in \textbf{Figure~\ref{fig:Fig1}(d)}, revealing that calcium exhibits the highest intensity counts due to the vaterite matrix, while iron and oxygen show lower intensity counts. Notably, the iron signal displays a prominent peak near the surface, indicating that the Fe$_3$O$_4$ MNPs are primarily embedded closer to the microsphere’s outer region. \textbf{Figure~\ref{fig:Fig1}(e)} shows a scanning electron microscopy (SEM) image of an individual magnetic vaterite microsphere, highlighting its porous surface morphology. These microspheres are expected to possess channel-like internal pore structures with sizes ranging from 30 to 50~nm, as supported by previous studies~\cite{Sukhorukov2004}. The synthesized vaterite particles are predominantly spherical, though some exhibit slight elongation, with diameters ranging from $2~\upmu$m, as observed in both the SEM image (Figure~\ref{fig:Fig1}(e)) and the optical microscopy image of vaterite microspheres drop-casted onto a glass coverslip (\textbf{Figure~\ref{fig:Fig1}(f)}).

\threesubsection{Quantum diamond microscopy}

The custom-built widefield QDM is based on the design detailed in Ref.~\cite{Fescenko2019}, with additional improvements for rapid inline data processing~\cite{Fescenko2023} and automated control~\cite{Berzins2022b} of an external magnetic field capable of reaching up to 222~mT. For convenience, we provide a concise overview of the measurement setup and highlight the modifications implemented to study magnetic microspheres. 

The sensor employed in this study is fabricated from a high-pressure, high-temperature type-Ib monocrystalline diamond (Sumitomo Electric) with dimensions of $2\times 2\times 0.06$~mm$^3$, grown along the (110) crystal plane and featuring an initial nitrogen concentration of $\sim$100~ppm. NV centers were created by implanting $^4$He$^+$ ions at a dose of $10^{12}$~He$^+/$cm$^2$ and energies of 5~keV, 15~keV, and 33~keV to achieve a uniform vertical vacancy distribution. After implantation, the diamond underwent a two-step annealing process: first at 800$^{\circ}$C for 4~hours, followed by a second step at 1100$^{\circ}$C for an additional 4 hours. This treatment facilitated the intracrystalline migration of vacancies and the formation of NV centers. This procedure is described in detail in Ref.~\cite{Berzins2022b} for F1 sample,  resulting in an NV center concentration of $\sim$6 ppm and the formation of a $\sim$200~nm NV layer near the diamond surface.

\begin{figure*}[!t]
  \begin{center}
  \includegraphics[width=0.9\linewidth]{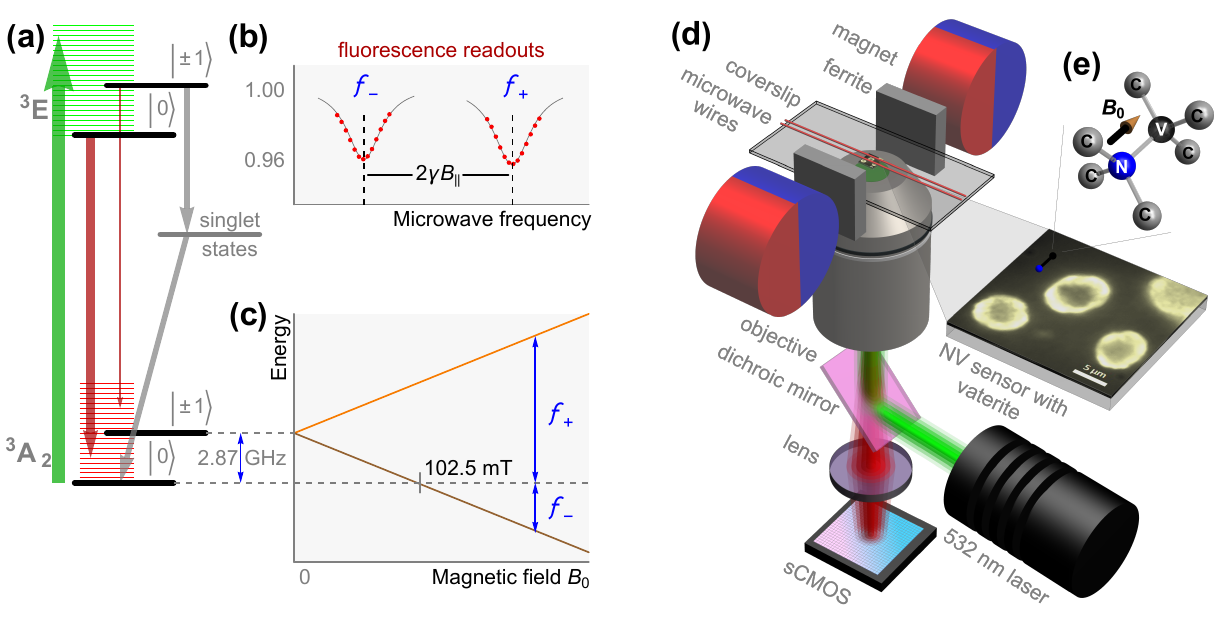}
  \end{center}
  \caption{\textbf{Quantum diamond microscope (QDM) for magnetic imaging using NV centers.} (a) Energy level diagram of the NV center in diamond, showing optical excitation (green arrow), spin-dependent fluorescence (red), and non-radiative decay via singlet states (gray). (b) Optically detected magnetic resonance (ODMR) spectrum, where fluorescence readout reveals resonance dips at frequencies $f_+$ and $f_{-}$ split by the magnetic field. (c) Zeeman splitting of the NV center ground state as a function of the applied magnetic field ($B_0$). (d) Schematic of the QDM, including magnets, microwave wires, an optical detection system, and an NV-diamond sensor with vaterite microspheres. (e) Atomic structure of the NV center and an optical image of the NV sensor detecting vaterite microspheres.}
  \label{fig:Fig2}
\end{figure*}

\textbf{Figures~\ref{fig:Fig2}(a-c)} illustrate the principles of the ODMR technique utilizing negatively charged NV centers in diamond (\textbf{Figure~\ref{fig:Fig2}(e)}). The electronic energy levels of the negatively charged NV center consist of a ground state ($^3$A$_2$) and an excited state ($^3$E), both of which are spin-triplet states (Figure~\ref{fig:Fig2}(a)). The $^3$A$_2$ ground state exhibits a zero-field splitting of $D \approx 2.87$~GHz between the $m_s = |0\rangle$ and $m_s=|\pm1\rangle$ spin sublevels, arising from electron spin-spin interactions. When the diamond is illuminated with 532~nm green light, electrons in the ground state are excited through vibronic levels of the lattice to the $^3$E state while preserving their spin projections. The excited triplet system can relax either radiatively ($^3$E → $^3$A$_2$), emitting fluorescence in the range of $600-800$~nm, or nonradiatively via intermediate singlet states. Under continuous green light excitation, the unequal probabilities of nonradiative relaxation from different excited sublevels lead to optical spin polarization into the $m_s = |0\rangle$ ground state sublevel. Electrons polarized in this state are more likely to undergo radiative relaxation, contributing maximally to the fluorescence intensity. However, when a microwave (MW) field with a frequency resonant to the $m_s = |0\rangle \leftrightarrow m_s = |\pm1\rangle$ ground-state transition is applied, a significant portion of the spin population is pumped into the $m_s = |\pm1\rangle$ sublevels. This results in a switch to nonradiative relaxation pathways under continuous green light excitation, leading to a decrease in fluorescence intensity. By sweeping the MW frequency around the resonant frequency, Lorentzian-shaped dips in the fluorescence intensity are observed (Figure~\ref{fig:Fig2}(b)). Applying a bias magnetic field $B_0$ lifts the degeneracy of the $m_s = |\pm1\rangle$ states due to the Zeeman effect. For a magnetic field $B_0 = B_{||}$ aligned along one of the NV axis orientations (e.g., the [111] crystallographic axis of the diamond), the transition frequencies $f_+$ and $f_-$ of the $m_s = |0\rangle \leftrightarrow m_s = |\pm1\rangle$ states shift linearly with the magnetic field (Figure~\ref{fig:Fig2}(c)). These frequencies are given by  $f_{\pm} = D \pm \gamma B_{||}$, where $\gamma \approx 28$~GHz/T is the electron gyromagnetic ratio. Since the zero-field splitting $D$ is sensitive to temperature and strain~\cite{Berzins2023}, it is eliminated by measuring the frequency difference $f_+ - f_- = 2\gamma B_{||}$, which provides an absolute measure of the magnetic field $B_{||}$. For magnetic fields $B_{||} > 102.5$~mT, the ground-state spin level anti-crossing~\cite{Auzinsh2019} must be considered, and the equation $2\gamma B_{||} = f_+ + f_-$  is used instead (Figure~\ref{fig:Fig2}(c)). Thus, magnetic field $B_{||}$ is optically detected using the ODMR technique by observing the corresponding shifts in the fluorescence intensity dips (Figure~\ref{fig:Fig2}(b)).

The QDM is operated on a standard vibration isolation optical table in a climate-controlled laboratory environment. \textbf{Figure~\ref{fig:Fig2}(d)} shows a schematic representation of the epifluorescence QDM setup. MW frequencies $f_+$ and $f_-$ are generated using a Stanford Research Systems SG384 frequency generator and delivered via two parallel copper wires (diameter = 0.11 mm). The studied Fe$_3$O$_4$-vaterite microspheres are positioned in a $\sim0.3$~mm wide gap between these wires. The diamond sensor is pressed against a glass coverslip (thickness = $100~\upmu$m) by the MW wires. The coverslip is mounted on a three-axis motorized stage (Newport, Picomotor 8742), which is controlled through a USB interface. A pair of diametrically magnetized permanent magnet cylinders, equipped with electromechanical rotation control and ferrite flux homogenizers~\cite{Berzins2022b}, is used to apply a magnetic field $B_{||} = 222$~mT. The NV centers in the diamond sensor are continuously excited over an area of  $\sim30 \times 30~\upmu$m$^2$ (top surface of the sensor) through the coverslip using a 532~nm green laser (Sprout-G-10W, Lighthouse Photonics Inc.) with a power of $\sim200$~mW. The laser light is focused through a $100\times$ microscope objective with a numerical aperture of 1.25 ($100\times$/1.25 Oil, $\infty$/0.17, ZEISS). The emitted red fluorescence is collected by the same objective and spectrally separated using a dichroic mirror (Thorlabs DMLP567R) and a long-pass filter (Thorlabs FEL0600). The fluorescence is then delivered via a tube lens (Thorlabs TTL200) to either an sCMOS sensor of an Andor Neo~5.5 camera or an avalanche photodiode detector (Thorlabs APD410A). The photodiode detector is connected to a digital oscilloscope (Yokogawa DL6154) for faster tuning and alignment (trimming) of the NV axis along the magnetic field $B_0$. The same QDM setup can also be used to obtain optical images by collecting white light delivered through the same objective and reflected from the sensor surface and the magnetic vaterite microspheres on it.

The MW sweeps are triggered by a pulse from the sCMOS camera, which records twelve $400 \times 400$ pixel frames per sweep. Twenty series of frames are accumulated and processed using a LabVIEW virtual interface, where they are converted into a single magnetic image of $B_{\text{str}}$ — the additional magnetic field component along the NV axis due to the sample — over an area of $26 \times 26~\upmu$m using a weighted average method~\cite{Fescenko2023}. The magnetic images are accumulated over $\sim 5$ minutes to reduce noise, while fast inline processing minimizes the impact of ODMR drift during accumulation. The saved magnetic images are further analyzed using Wolfram Mathematica.

\section{Results and discussion}
\begin{figure*}[!t]
  \begin{center}
  \includegraphics[width=0.9\linewidth]{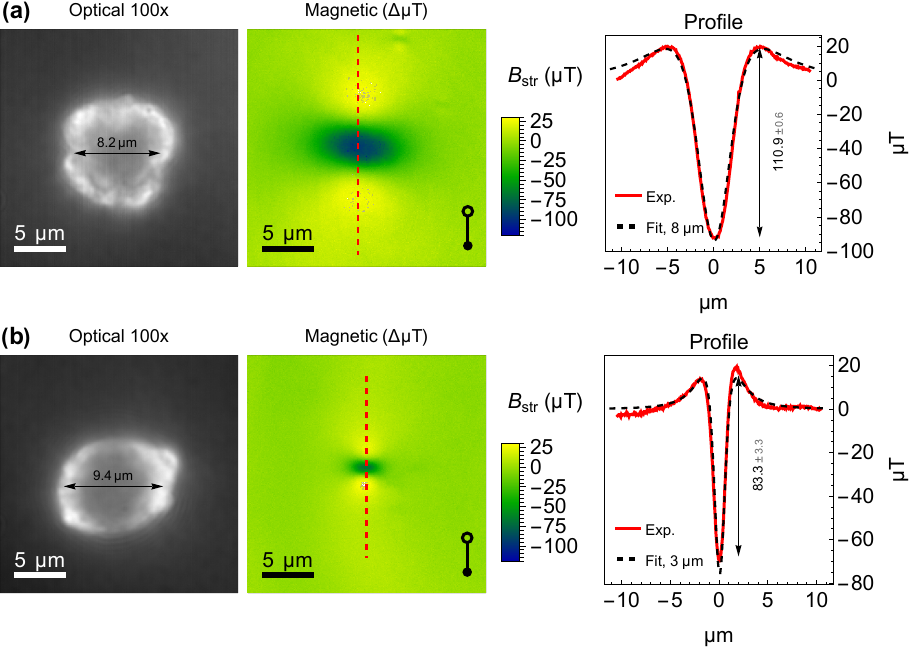}
  \end{center}
  \caption{\textbf{Optical and magnetic imaging of vaterite microspheres with quantum diamond microscopy.} (a) Optical (left) and magnetic (middle) images of a vaterite microsphere uniformly functionalized with 20-nm Fe$_3$O$_4$ magnetic nanoparticles. The corresponding magnetic field profile (right) along the red dashed line reveals a symmetric dipole-like signal with a PP amplitude of 110.9 ± $0.6~\upmu$T. (b) Optical and magnetic images of a vaterite microsphere exhibiting non-uniform magnetization, functionalized with 10-nm Fe$_3$O$_4$ MNPs. The magnetic profile shows a weaker and more localized field distribution with a PP amplitude of 83.3 ± $3.3~\upmu$T. Black dashed lines are simulated profiles for uniformly magnetized spheres with diameters of 8 and $3~\upmu$m, correspondingly. }
  \label{fig:Fig3}
\end{figure*}

In three samples of vaterite microspheres containing Fe$_3$O$_4$ MNPs (diameters of 5~nm, 10~nm, and 20~nm), we identified $>35$ individual vaterite particles using reflective optical imaging under white light illumination. All microspheres exhibit nearly spherical shapes with diameters of several micrometers, while the depth of field of the microscope objective is approximately one micrometer. To assess particle size, we adjusted the image plane to the center of the microspheres, maximizing the apparent size of the particle. This approach yields a sharp image of the surface along the sphere’s great circle, while the rest of the sphere’s surface appears blurred and dark. Two representative vaterite microspheres are shown in \textbf{Figure~\ref{fig:Fig3}}, left panels. Throughout the subsequent text, we designate them as Microsphere A (Figure~\ref{fig:Fig3}(a)) and Microsphere B (Figure~\ref{fig:Fig3}(b)). 

For magnetic imaging, we refocused the system onto the diamond surface containing the NV centers. Magnetic images of the corresponding microspheres are presented in the middle panels of Figure~\ref{fig:Fig3}. The dumbbell legend, featuring open and closed circles, indicates the alignment of the NV centers relative to the image plane, and thus the direction of the applied magnetic field ($B_{\text{app}}$), and the measured component of the stray field ($B_{\text{str}}$). Despite the microspheres’ size of several micrometers, the shapes of the magnetic signals closely resemble those of the point spread function of a magnetic dipole~\cite{Dolgovskiy2016}, with magnetization aligned along the measured component of the magnetic field. This magnetic pattern shows two faint yellow maxima of field strength and a prominent blue minimum between them, set against a green zero-field background.

We further reduced the dimensionality by deriving magnetic profiles (Figure~\ref{fig:Fig3}, right panels) along a dashed red line connecting the two maxima within the 2D magnetic images. These measured magnetic profiles (red solid lines) were compared with simulated profiles (black dashed lines) generated \textit{via} finite-element analysis, modeling uniformly magnetized spheres of varying diameters. The simulations reveal that the width of the dip in the profile depends on the diameter of the magnetized particle, enabling us to determine the “magnetization diameter” of each measured microsphere by fitting the measured and simulated profiles. For example, Microsphere A, with a diameter of 8.2~$\upmu$m, is accurately represented by an 8-$\upmu$m “magnetization diameter.” We found that this “magnetization diameter” does not always correspond to the actual diameters of the microspheres in the studied samples. For instance, Microsphere B has a diameter of 9.4~$\upmu$m, yet its magnetic signal matches that of a 3-$\upmu$m sphere, as evidenced by the fit in Figure~\ref{fig:Fig3}(b). We hypothesize that Microsphere B exhibits reduced porosity across most of its surface, with MNPs penetrating only one side, compactly occupying a volume equivalent to that of a 3-$\upmu$m sphere. The EDS map in Figure~\ref{fig:Fig1}(c) reveals a similar uneven deep penetration of MNPs, particularly in the lower-left region of the microparticle. Note that the magnetic pattern approach can confirm whether a sphere is uniformly magnetized along all angles in spherical coordinates, but it is insensitive to the radial distribution of MNPs. Specifically, a vaterite sphere with a surface uniformly functionalized with MNPs produces the same magnetic signal as an identical sphere with the same total amount of MNPs uniformly distributed throughout its volume.

The magnetic profiles are utilized to assess the PP amplitude, as illustrated in the right panels of Figure~\ref{fig:Fig3}. This amplitude serves as a direct measure of the magnetic contrast, which is of interest for the biomedical applications of such composite particles. Furthermore, the PP amplitude of the simulated profiles is unequivocally correlated with the volume magnetization, serving as a fitting parameter. By fitting the measured magnetic profile we can estimate the number ($N$) of MNPs captured by the vaterite microsphere and contributing to the magnetic signal. Using the volume $V_{\text{np}}$ of the Fe$_3$O$_4$ MNPs and their magnetization $M_{\text{np}}$ from specifications, along with the volume $V_{\text{ms}}$ and magnetization $M_{\text{ms}}$ of the microspheres derived from the fit, we calculate the number of captured MNPs as $N = (V_{\text{ms}}\cdot M_{\text{ms}}) / (V_{\text{np}}\cdot M_{\text{np}})$. For Microsphere A, with a diameter of 8~$\upmu$m, the magnetization $M_{\text{ms}} = 285$~A/m. It contains 20-nm Fe$_3$O$_4$ MNPs with a magnetization $M_{\text{np}} = 3.89-4.14\times10^6$~A/m. This yields N = 4412 ± 391 MNPs captured by Microsphere A. Similarly, for Microsphere B, with a magnetization $M_{\text{ms}}= 550$~A/m and a 3-$\upmu$m 'magnetization diameter,' the number of 10-nm Fe$_3$O$_4$ MNPs with $M_{\text{np}}= 3.63-4.14\times10^6$~A/m is $N = 3839 \pm 356$. Thus, this estimate confirms the capacity of a single vaterite microsphere to concentrate and retain thousands of MNPs. 

Furthermore, the magnetic image in Figure~\ref{fig:Fig3}(a) exhibits distinctive magnetic signals in the top-right corner, originating from two small MNPs. These are likely individual Fe$_3$O$_4$ MNP that have detached from the vaterite microsphere. Their magnetic signals are relatively strong due to their proximity to the NV layer. The sharp magnetic signal from a single 20-nm Fe$_3$O$_4$ MNP is averaged over an optical resolution area of approximately $0.5 \times 0.5$~$\upmu$m$^2$.

\begin{figure*}[!t]
  \begin{center}
  \includegraphics[width=0.9\linewidth]{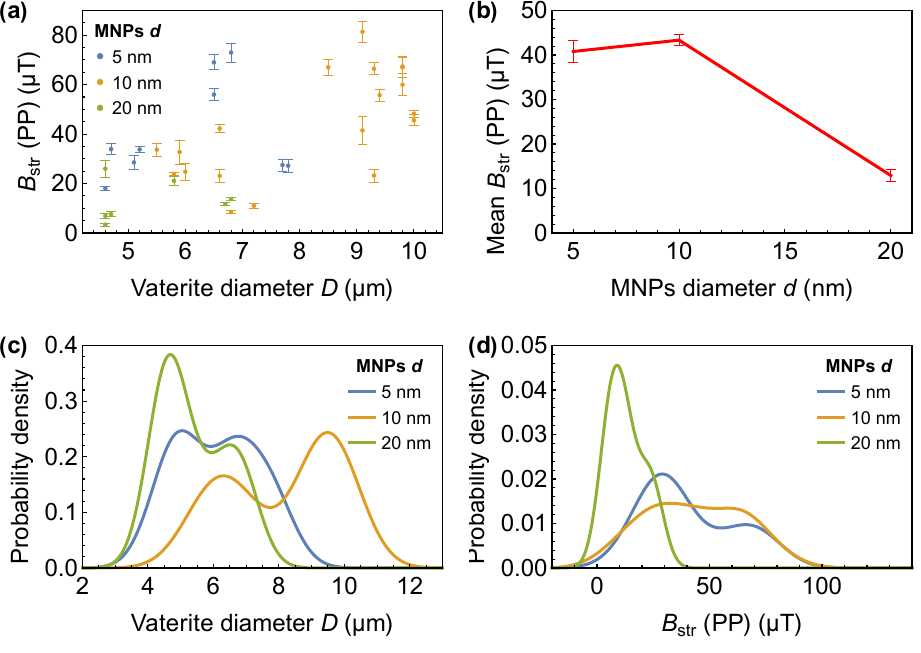}
  \end{center}
  \caption{\textbf{Stray magnetic fields of vaterite microspheres with embedded Fe$_3$O$_4$ MNPs.} (a) Stray field measured below vaterite microspheres as a function of their diameter, for different Fe$_3$O$_4$ NP sizes (5-nm, 10-nm, and 20-nm). PP amplitude of the stray magnetic field. (b) Mean stray field as a function of Fe$_3$O$_4$ MNP diameter, showing a decrease for larger Fe$_3$O$_4$ size. (c) Probability density distribution of vaterite microsphere diameters for different Fe$_3$O$_4$ MNP sizes. (d) Probability density distribution of stray field values, illustrating the dependence of magnetic signal strength on Fe$_3$O$_4$ MNP size. }
  \label{fig:Fig4}
\end{figure*}

Finally, we applied the analysis described in Figure~\ref{fig:Fig3} to $> 35$ individual vaterite microspheres to investigate the magnetization uniformity of the prepared samples. The vaterite microspheres exhibited diameters ranging from 5~$\upmu$m to 10~$\upmu$m, with an average of $\sim7$~$\upmu$m. \textbf{Figure~\ref{fig:Fig4}(a)} presents the PP amplitudes of the stray magnetic field as a function of microsphere diameter. These PP amplitudes range from 3.4 to 81~$\upmu$T, excluding several outliers not displayed on the plot. \textbf{Figure~\ref{fig:Fig4}(b)} illustrates the PP amplitudes averaged across the samples. Samples containing 5-nm and 10-nm superparamagnetic Fe$_3$O$_4$ MNPs exhibit significantly higher stray field amplitudes ($41 \pm 1$~$\upmu$T) compared to those with 20-nm ferri-MNPs ($12 \pm 1$~$\upmu$T). This reduction in magnetic field amplitude may result from the mutual cancellation of ferromagnetic domains or may be attributed to the larger diameter of the 20-nm Fe$_3$O$_4$ MNPs, which could limit their ability to penetrate the vaterite pores. The last hypothesis is supported by the observation that Microsphere B, as shown in Figure~\ref{fig:Fig3}, contains over 3,800 nanoparticles with a diameter of 10~nm within a 3-$\upmu$m 'magnetization diameter,' whereas Microsphere A contains more than 4,400 nanoparticles with a diameter of 20~nm, despite having a volume approximately 19 times larger.

These data are presented as probability density distributions of microsphere diameters (\textbf{Figure~\ref{fig:Fig4}(c)}) and PP amplitudes (\textbf{Figure~\ref{fig:Fig4}(d)}). Although the vaterite microspheres of the sample with 20-nm Fe$_3$O$_4$ MNPs exhibit a size distribution similar to that of the sample with 10-nm Fe$_3$O$_4$ MNPs, their PP amplitudes are markedly distinct from those of samples with $5-10$~nm MNPs. This step-like change in stray field amplitude more strongly supports a phase transition from superparamagnetism to ferrimagnetism than a gradual decrease in NP penetration into the vaterite matrix. 
We now evaluate the potential of these MNP-loaded microspheres as $T_2$ contrast agents in magnetic resonance imaging (MRI). A 40~$\upmu$T signal, measured at a distance of several micrometers from a microsphere containing approximately 4,000 MNPs, indicates a significant local effect. While this field is not substantial for a single particle, its scalability suggests promising applications. For comparison, standard Fe$_3$O$_4$-based contrast agents, such as Feridex, generate fields of $10-100$~$\upmu$T at distances of a few nanometers. A local field of 40~$\upmu$T can substantially shorten $T_2$ relaxation times by inducing magnetic field gradients across a region spanning several micrometers. The $T_2$-shortening effect scales with the square of the magnetic field ($\Delta B^2$) and diminishes inversely with the sixth power of the distance ($1/r^6$). For a single 20-nm particle, this field decays rapidly; however, aggregation or higher concentrations amplify the effect. At a concentration of approximately 0.1~mM, corresponding to hundreds of such microspheres per milliliter of tissue, the $T_2$ relaxation time could be reduced by 20–50\%, which is sufficient to produce a noticeable contrast in MRI. The vaterite loading process relies on diffusion into the mesoporous vaterite host and can be further enhanced using cold infusion methods~\cite{Noskov2021} if higher concentrations are required.

\section{Conclusion}

We loaded vaterite microspheres with Fe$_3$O$_4$ MNPs and characterized the accumulated stray magnetic fields from individual vaterite microspheres using a widefield optical microscopy technique. The observed magnetic patterns align with theoretical models assuming uniform magnetization of the vaterite microspheres. The PP amplitudes of the magnetic signals suggest that each vaterite microsphere contains several thousand embedded MNPs. Additionally, we identified a phase transition from superparamagnetic to ferromagnetic behavior between 10-nm and 20-nm MNPs, manifested as a significant reduction in total magnetization. While this paper does not address these aspects, future studies will investigate the impact of varying vaterite microsphere sizes, shapes and the controlled incorporation of magnetic materials. These efforts are expected to enhance the efficiency and precision of drug loading and release, thereby facilitating more effective drug delivery.

Our method offers new insights into the magnetic behavior of MNPs by enabling the observation of magnetism in individual micro- and nanoparticles. A primary limitation is the optical resolution, which prevents the resolution of individual MNPs in samples with a density exceeding one particle per $0.5 \times 0.5$~$\upmu$m$^2$ area. However, at low densities, even a single MNP can be detected owing to the high sensitivity of quantum diamond imaging. The NV microscope employed in this study can generate magnetic images of features as small as 1.4~$\upmu$T within less than five minutes, utilizing parallel ODMR detection across a $400 \times 400$ pixel array. Further refinements and optimization could potentially enable the attainment of the predicted nanotesla sensitivity.

\medskip

\medskip
\textbf{Acknowledgements} \par 
This work was supported by the “Latvian Quantum Technologies Initiative” project under the European Union Recovery and Resilience Facility, Grant No. 2.3.1.1.i.0/1/22/I/CFLA/001. Mona Jani would like to acknowledge the financial assistance from the FORTHEM Alliance for secondment research stays. Pavel Ginzburg acknowledges the funding support from the Israel Science Foundation (ISF grant number 1115/ 23), ERC StG “In Motion” (Grant No. 802279), and the Ministry of Science, Technology, and Space of Israel (Grant No. 79518). 

\medskip

%

\bibliographystyle{MSP}
\bibliography{main}


\end{document}